\documentclass[prb,twocolumn,preprintnumbers,amsmath,amssymb,floatfix]{revtex4-1}

\usepackage{graphicx}
\usepackage{amssymb}
\usepackage{amsmath}
\usepackage{dcolumn}
\usepackage{bm}
\usepackage{color}

\makeatletter 
\def\@hangfrom@section#1#2#3{\@hangfrom{#1#2#3}}
\makeatother

\makeatletter        
\def\@biblabel#1{[#1]}
\makeatother

\newcommand{\cmt}[1]{} 
\newcommand{\hlt}[1]{#1} 

\usepackage[normalem]{ulem}
\usepackage{xcolor}
\renewcommand\sout{\bgroup\markoverwith{\textcolor{red}{\rule[0.5ex]{2pt}{0.4pt}}}\ULon}

\mathchardef\mhyphen="2D

\usepackage{hyperref}
\begin{document}

\title{\hlt{Is hydrogen diffusion in amorphous metals non-Arrhenian?}}

\author{Chunguang Tang$^{1,2}$}
\email{chunguang.tang@anu.edu.au}
\author{Gang Sun$^3$}
\author{Yun Liu$^{1}$}
\email{yun.liu@anu.edu.au}

\affiliation{$^1$ Research School of Chemistry, The Australian National University, Canberra, Australia; \\$^2$ Institute of Climate, Energy and Disaster Solutions, The Australian National University, Canberra, Australia; \\$^3$ Department of Fundamental Engineering, Institute of Industrial Science, University of Tokyo, 4-6-1 Komaba, Meguro-ku, Tokyo 153-8505, Japan.}

\begin{abstract}
Hydrogen diffusion is critical to the performance of metals for hydrogen storage as well as other important applications. As compared to its crystalline counterpart which follows the Arrhenius relation, hydrogen diffusion in amorphous metals sometimes are experimentally found to be non-Arrhenian. In this work we studied the diffusion of hydrogen in amorphous Pd-H and Zr-Cu-H alloys based on molecular dynamics simulations. Our simulations confirm Arrhenian diffusion behaviour for hydrogen in amorphous alloys, in contrast to previous theoretical studies which predict non-Arrhenian behaviour. We show that the simulated non-Arrhenian diffusion based on molecular dynamics could result from a systematic error related to too short simulation time. We also discussed the experimental non-Arrhenian behaviour of hydrogen diffusion within the framework of quantum tunneling and amorphous-amorphous phase transformations.

 \vspace{10pt}

Published at \href{https://doi.org/10.1016/j.ijhydene.2022.01.023}{International Journal of Hydrogen Energy, 47, 9627 (2022).}

\end{abstract}
\maketitle
\section{Introduction}
Hydrogen, as the smallest element, can easily diffuse into and interact with materials. Specifically, the interactions between hydrogen and metals have remained as an important research topic for more than a centurary. For example, the old phenomenon of hydrogen embrittlement \cite{johnson_remarkable_1875, zhong_computer-simulation_1993, song_atomic_2013} is a factor to be considered for the design of internal combustion engines for hydrogen-rich fuels \cite{saborio_gonzalez_review_2018} and transport of hydrogen through pipelines \cite{takayama_hydrogen_2011, thomas_hydrogen_2020}. Recently metal hydrides have attracted extensive research activities for their broad applications, such as hydrogen storage \cite{rusman_review_2016}, smart window and sensors \cite{yoshimura_metal_2013}. Specifically, the past decades have seen intensive research interests in amorphous metals for hydrogen-related applications \cite{yamaura_hydrogen_2005, dolan_composition_2006,lin_hydrogenation_2019}. Compared to their crystalline counterparts, amorphous metal hydrides exhibit some favorable properties, such as higher hydrogen storage capacity \cite{maeland_hydrides_1980, lin_room_2012}, easy tuning of dehydrogenation conditions \cite{lin_towards_2016}, and enhanced (de)hydrogenation kinetics of metal hydrides \cite{zhang_remarkably_2015, el-eskandarany_metallic_2019}.

In the above examples and also a wider range of applications \cite{gapontsev_hydrogen_2003}, hydrogen diffusion is a critical factor that impacts on material properties and performance. The diffusion of hydrogen in crystalline metals is relatively simple and the temperature ($T$) dependence of diffusivity $D$ follows the Arrhenius equation
\begin{equation}
D=D_0 {\rm{exp}}(\frac{-E_a}{RT})
\label{eq:D}
\end{equation}
where $E_a$ is the activation enthalpy for the diffusion, $R$ is the gas constant, and $D_0$ is a constant. On the other hand, hydrogen diffusion in amorphous metals exhibits complicated features. For example, it is sensitive to hydrogen concentrations, which can be explained by the assumption of Gaussian energy distribution for hydrogen in various interstitial sites \cite{kirchheim_hydrogen_1988}. At higher hydrogen concentrations, hydrogen atoms tend to occupy those higher-energy sites, resulting in increased diffusivity. Hydrogen diffusion in amorphous metals intriguingly deviate from the Arrhenius equation. For example, in a series of amorphous metals \cite{bowman_nmr-studies_1981, bowman_proton_1982, bowman_hydrogen_1983}, such as TiCuH$_x$ ($x$ from 1.3 to 1.4) and Zr$_2$PdH$_{2.88}$, the Arrhenius plot of log$D$ versus $1/T$ breaks into several segments over the temperature range between 115 K and 420 K, with the activation enthalpy being higher at higher temperatures. \hlt{Although the normal Arrhenius relation is also observed in many amorphous systems \cite{lee_hydrogen_1985, markert_hydrogen_1988}, the temperature ranges for these measurements are small compared to those of the non-Arrhenian datasets, and it is not clear how the diffusion behaviour changes in a larger temperature range.}

The non-Arrhenian hydrogen diffusion has attracted a number of theoretical investigations. Monte Carlo simulations \cite{kirchheim_monte-carlo_1987} have reported negative curvatures of the Arrhenius curve (i.e., $E_a$ decreases as $T$ increases) or linear Arrhenius lines, depending on the assumption of a Gaussian distribution or a single value of hydrogen interstitial site energy. A structure model that relates the negative curvature to the temperature dependence of the short-range order in amorphous metals has also been proposed \cite{eliaz_new_1999}. On the other hand, a molecular dynamics simulation work \cite{lee_comparative_2014} has reported a positive curvature for H diffusion in Zr-Cu alloys. Overall, however, these theoretical studies (for the Monte Carlo simulations \cite{kirchheim_monte-carlo_1987}, if Gaussian distribution is assumed) predict rather smoothly curved Arrhenius plots, instead of linear segments as observed in experiments. A recent study based on Monte Carlo simulation \cite{hung_estimation_2010} reported a kinked Arrhenius relation for hydrogen diffusion in amorphous iron, but without a detailed discussion of it. 

\hlt{As can be seen, to date the diffusion of hydrogen in amorphous metals is not well understood. The theoretical interpretations of hydrogen diffusion in amorphous metals are inconsistent with each other as well as inconsistent with the experiments. Moreover, the mechanisms of the non-Arrhenian (or segmented Arrhenian) diffusion behaviour are not clear. There exists evidence that hydrogen also impacts on the structural/thermal stability and mechanical properties of amorphous alloys \cite{jayalakshmi_effect_2006}. Hence, for amorphous metals to reach their full potential for hydrogen-related applications, it is necessary for us to fundamentally understand the abnormal diffusion of hydrogen.} In an attempt to shed further light on this issue, we here study the diffusion of hydrogen in amorphous palladium (Pd) and ZrCu alloy based on molecular dynamics simulations. We choose these two systems partially because palladium, as a single-element metal, has very low glass forming ability while ZrCu is a good glass former and represents the prototype of many important amorphous alloys. As a first discovered metal with hydrogen absorption capacity \cite{graham_absorption_1866}, Pd still finds active applications in the hydrogen energy sector \cite{dekura_hydrogen_2019, lamb_ammonia_2019}.

\section{Methods}

\begin{figure}
\includegraphics[width=3.1in]{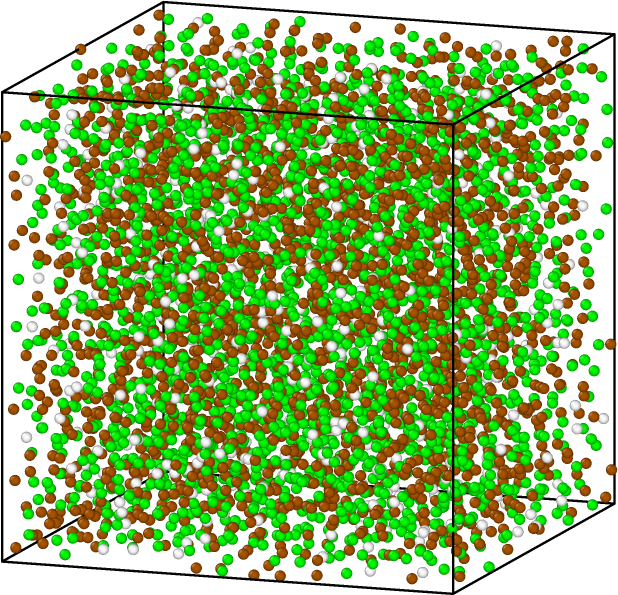}
\includegraphics[width=3.3in]{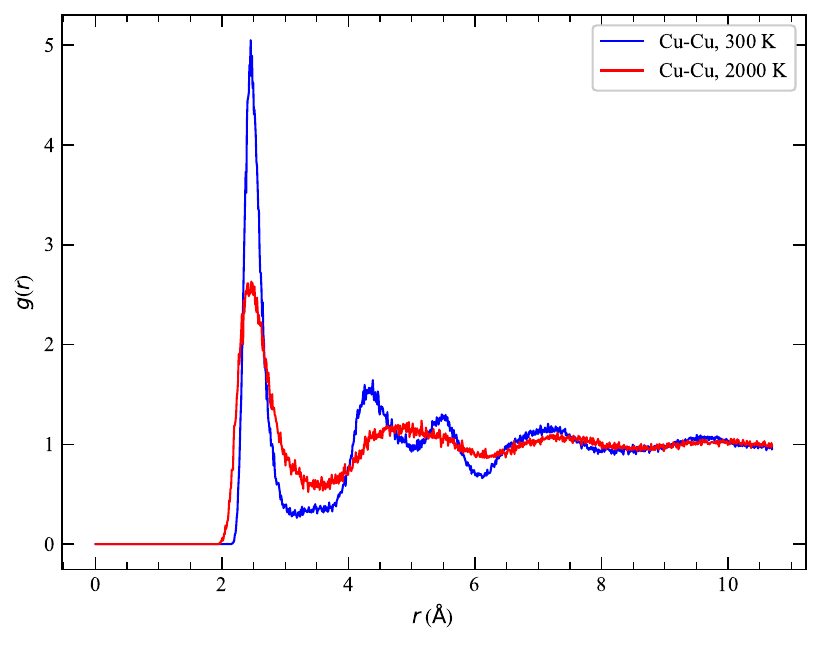}
\caption{Structure of Zr-Cu-H system. (top) Amorphous structure model at 300 K after melting-quenching. Cu: brown, Zr: green, H: white. (bottom) Cu-Cu pair distribution function of the Zr-Cu-H alloy at 2000 K (liquid state) and 300 K (glass state). The splitting of the second g(r) peak is characteristic of the amorphous structure. The functions for Zr-Zr pair and Pd-Pd pair for the Pd-H alloy are similar and not shown.}
\label{fig:rdf}
\end{figure}

To obtain the amorphous structures, the Pd-H alloy of composition H/Pd=5\% (containing 4000 Pd and 200 H atoms) was well liquefied at 2100 K and quenched to 300 K at 10$^{14}$ K/s, and the Zr-Cu-H alloy with H/metal=10\% (with 2197 Zr, 2197 Cu, and 439 H atoms) was well liquefied at 2000 K and quenched to 300 K at 1.25$\times$10$^{12}$ K/s. The amorphous states of the resulting structures were confirmed by their pair distribution functions (Fig. \ref{fig:rdf}). We annealed the amorphous structures at various temperatures to obtain the equilibrium system volume. These computations were carried out using the NPT (constant atom number, pressure, and temperature) ensemble by setting the normal pressures to zero. \hlt{Then for each temperature we built a structure by scaling the amorphous structure to the corresponding equilibrium volume and the mean squared displacements (MSDs) of H atoms were calculated using this structure based on the NVT (constant atom number, volume, and temperature) ensemble, following an initial relaxation (for 5 to 50 ps) that equilibrates the structure. For each temperature we performed 10 sets of MSD computations by setting different initial velocities of the atoms. At low temperatures the initial relaxation upon different initial velocities may not result in very different structures of the metallic hosts because of the slow dynamics, and hence it remains unclear whether the resulting MSD data are representative. We note that, due to the enough large size of the systems, the single structure obtained from the melting-quenching is representative of the amorphous structure characteristics. To confirm this, we built 10 more independent Pd-H amorphous structures by further relaxing the system at 2100 K for 10 to 100 ps before quenching, and performed one MSD computation for each structure at 300 K following the above method. It was found that the MSD data based on these independent structures are essentially identical to those based on the single amorphous structure with different initial velocities.} 

The obtained MSDs can be used to compute the self diffusion coefficient of H atoms following Einstein relation
\begin{equation}
 D=\frac{1}{6}\lim_{t\to\infty}   \frac{\partial \langle r^2(t) \rangle}{\partial t} 
\label{eq:Einstein}
\end{equation}
where $t$ is time and $\langle r^2(t) \rangle$ is the ensemble average of the MSDs of diffusing atoms under investigation.  The atomic interactions of the Pd-H system were described by an embedded atom method (EAM) potential \cite{zhou_embedded-atom_2008}, and for the Zr-Cu-H system, a modified EAM potential \cite{lee_comparative_2014} was used. \hlt{Tests \cite{zhou_embedded-atom_2008, lee_comparative_2014} of thermal properties, hydrogen migration energies, and crystal lattice constants show comparable results with experiments and/or first principles computations, indicating the reliability of the potentials. } All the simulations were performed using the code LAMMPS \cite{plimpton_fast_1995} with Nos$\rm{\acute{e}}$-Hoover thermostat and barostat, and the timestep for all computations was set as 1 fs for Pd-H and 0.5 fs for Zr-Cu-H, respectively. \hlt{We chose 0.5 fs for Zr-Cu-H following ref \cite{lee_comparative_2014} for comparison.} Ten independent MSD calculations for each temperature were carried out for statistics unless otherwise specified.

\begin{figure}
\includegraphics[width=3.5in]{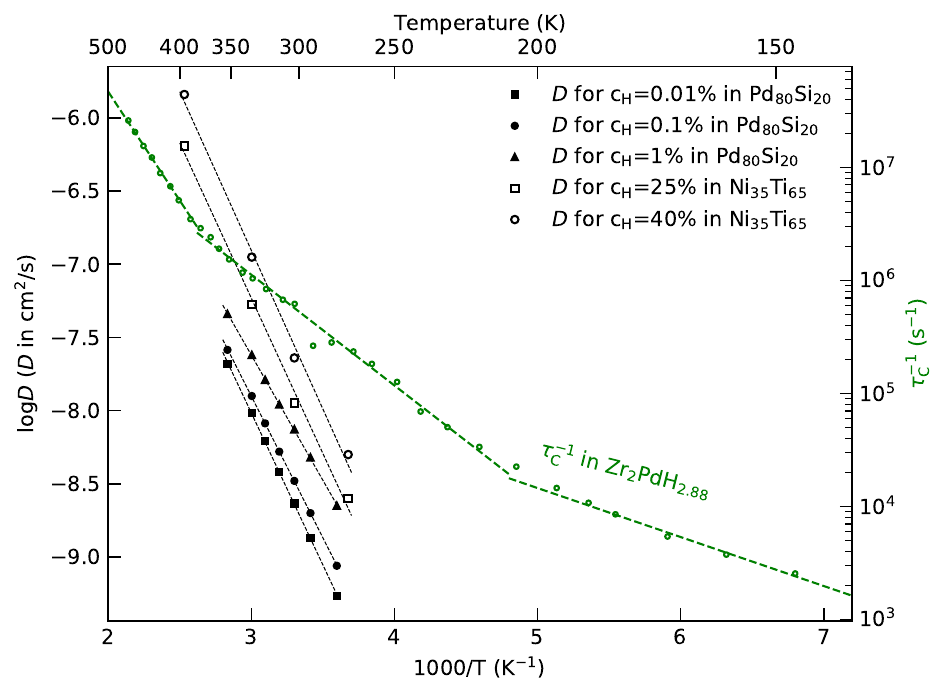}
\caption{Experimental hydrogen diffusion data replotted from Fig. 3 of reference [\citenum{kirchheim_modeling_1985}] for Pd$_{80}$Si$_{20}$ and Fig. 6 of [\citenum{kirchheim_hydrogen_1988}] for Ni$_{35}$Ti$_{65}$, with C$_{\rm{H}}$ being the ratio of hydrogen to metal. The inverse diffusion correlation time ($\tau_{\rm{C}}^{-1}$), based on proton rotating-frame relaxation and indicative of hydrogen diffusion coefficient\cite{bowman_nmr-studies_1981}, is replotted from reference [\citenum{bowman_hydrogen_1983}]. \hlt{The diffusion is Arrhenian in small temperature ranges but the Arrhenian behaviour becomes segmented in a larger temperature range.}}
\label{fig:expt}
\end{figure}

\section{Results and discussions}

\begin{figure*}
\includegraphics[width=6in]{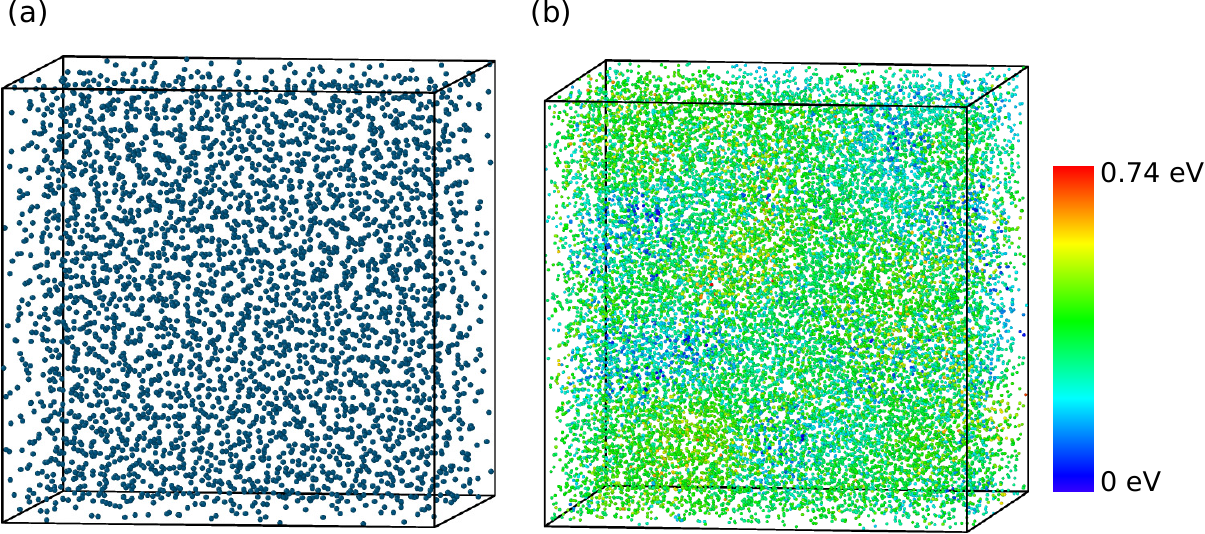}
\caption{(left) A fixed amorphous Pd host (4000 atoms) corresponding to 300 K, with each dot representing a Pd atom. (right) Relative potential energy of the system (4001 atoms) that contains a single test H atom at one of the $\sim$22300 Pd Voronoi vertices corresponding to the fixed Pd host. Each dot represents a vertex and its color represents the potential energy.}
\label{fig:space}
\end{figure*}

\begin{figure*}
\includegraphics[width=6in]{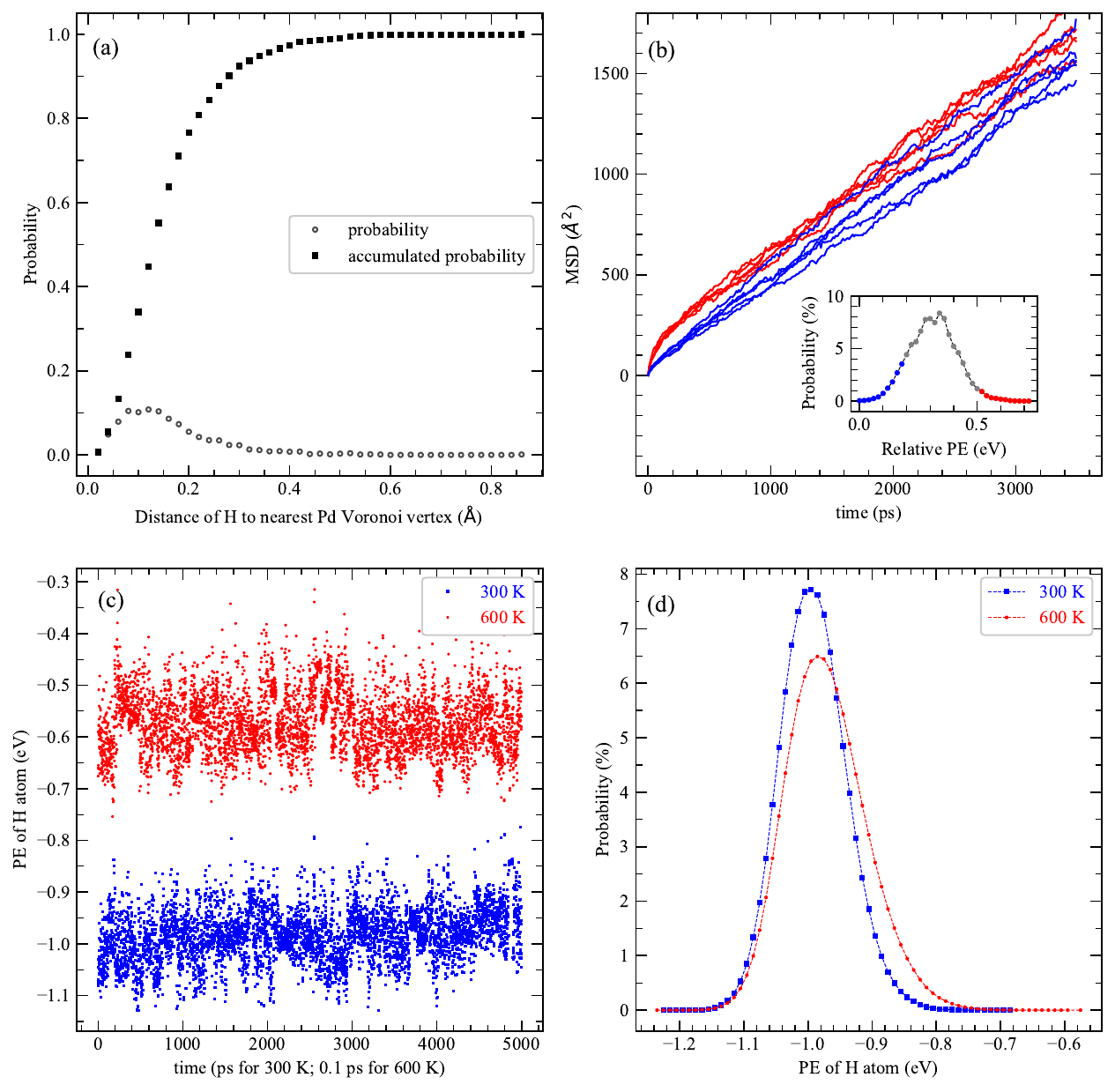}
\caption{H motion, at 300 K if not specified, in Pd-H alloys with Pd atoms fixed (see Fig. \ref{fig:space}(a)). (a) The distribution of hydrogen atoms. (b) Example mean squared displacement curves of H with high (red) and low (blue) initial potential energies (PEs). Inset: Relative system PE with only a test H atom at different vertices of Pd Voronoi polyhedra. (c) PE of an example atom as a function of time. The PE at 600 K is shifted upwards by 0.4 eV for clarity. (d) Temperature effect on PE distribution of H atoms. On average, the minimum PE of an H atom is -1.15$\pm$0.016 eV and -1.15$\pm$0.017 eV at 300 and 600 K, respectively.}
\label{fig:hmotion}
\end{figure*}

\begin{table}
\caption{Hydrogen diffusion data in amorphous Pd-H and Zr-Cu-H alloys. $t_0$ is the time span for computation, $\langle r^2 \rangle$ is the mean squared displacement value corresponding to time $t_0$, and $D$ value was obtained from the slope of linearly fitting $\langle r^2(t) \rangle$ versus $t$ using the top 50\% $\langle r^2(t) \rangle$ values. STD is for standard deviation.}\tabcolsep=1.5pt
\begin{tabular} {@{\extracolsep{8pt}}ccccc}  \hline  \hline
$T$ (K) & $t_0$ (ns) & $\langle r^2 \rangle$ ($ $\AA$^2$) &  $D$ (cm$^2$/s) & STD (cm$^2$/s) \\
\hline
\multicolumn{5}{c}{Pd-H}\\
125 & 0.1 & $\sim$11 &   1.48e-6  &  3.27e-7  \\
125 & 10 & $\sim$250 &   3.83e-7  &  8.00e-8  \\
150 & 0.1 & $\sim$20 &   2.80e-6  &  5.02e-7   \\
150 & 10 & $\sim$810 &   1.25e-6  &  1.59e-7   \\
200 & 0.1 & $\sim$60 &   9.80e-6  &  1.34e-6   \\
200 & 1 & $\sim$490 &   7.87e-6  &  8.70e-7   \\
250 & 0.1 & $\sim$140 &   2.17e-5  &  3.12e-6    \\
300 & 0.1 & $\sim$280 &   4.65e-5  &  4.80e-6    \\
350 & 0.1 & $\sim$440 &   7.14e-5  &  8.12e-6    \\
400 & 0.1 & $\sim$640 &   1.06e-4  &  1.19e-5    \\
450 & 0.1 & $\sim$820  &   1.39e-4  &  1.53e-5    \\
500 & 0.1 & $\sim$1110  &   1.82e-4  &  1.66e-5    \\
\hline
\multicolumn{5}{c}{Zr-Cu-H}\\
400 & 0.1 & $\sim$1.3  &   1.30e-7  &  3.80e-8    \\
400 & 10 & $\sim$22  &   2.88e-8  &  3.67e-9    \\
500 & 0.1 & $\sim$3.1  &   3.84e-7  &  9.20e-8    \\
600 & 0.1 & $\sim$7.2  &   9.62e-7  &  1.85e-7    \\
700 & 0.1 & $\sim$15  &   2.26e-6  &  3.80e-7    \\
800 & 0.05 & $\sim$17  &   5.13e-6  &  9.14e-7    \\
1000 & 0.05 & $\sim$60  &   1.81e-5  &  1.63e-6    \\
1100 & 0.05 & $\sim$90  &   2.92e-5  &  2.55e-6    \\
\hline\hline
\end{tabular}
\label{tab:d}
\end{table}

We start our study by investigating the local dynamics of hydrogen since, based on previous theoretical studies, the diverse local environment of hydrogen interstitials impacts on its diffusion behaviour. Understanding the detailed local dynamics in amorphous materials represents a nontrivial task and here we only study the site energetics of H atoms with Pd atoms fixed for simplification. \hlt{The fixed Pd atoms are shown in Fig. \ref{fig:space}(a).} At 300 K, we collected H positions at the end of MSD calculations and geometrically optimized their positions. Fig. \ref{fig:hmotion}(a) shows the distribution of their distance to the nearest vertex of Pd Voronoi polyhedra (i.e., Pd Wigner–Seitz cells), with $\sim$90\% of the atoms coincide with the vertices in a tolerance of 0.3 $ $\AA$ $. This indicates that the Voronoi vertices are preferred interstitial sites, although there may be a small fraction of unfavorable vertices corresponding to small open volume \cite{lancon_simulation_1985}. We then put a test H atom at various vertices and computed the potential energy of the system without geometrical optimization. \hlt{Fig. \ref{fig:space}(b) shows the space distribution of the relative potential energy of the single H atom. The statistics of the relative potential energy features a bell shape, as shown in the inset of Fig. \ref{fig:hmotion}(b). We performed five sets of MSD calculations to test the effects of high and low, respectively, site energy on diffusion. For each calculation, we placed 200 H atoms on randomly selected low (or high) energy vertices, with at least 2.5 $ $\AA $ $ apart from each other, and computed their MSDs after optimizing their initial positions.} As indicated in Fig. \ref{fig:hmotion}(b), initially the atoms of high energies move much faster than those of low energies, but after $\sim$100 ps the two groups of MSD curves exhibit similar slopes. This implies that although local environment impacts on the movement of H atoms, the diffusion coefficient, which is a long range property, reflects the features of the global potential energy surface of the system.

To address the temperature effect, we relaxed H atoms at 300 and 600 K, respectively, with Pd atoms fixed. Fig. \ref{fig:hmotion}(c) shows the potential energy of an example atom as a function of the relaxation time, and Fig. \ref{fig:hmotion}(d) shows the potential energy distribution of all H atoms over the span of relaxation. As can be seen, as the temperature increases, the probability of H atoms in high energy states increases, but their preference for low-energy sites does not change. Indeed, we calculated the average lowest potential energy of the atoms to be the same for 300 and 600 K. For the effect of temperature on diffusion in amorphous materials, some Monte Carlo simulations\cite{kirchheim_monte-carlo_1987,limoge_temperature_1990} assume the increasing probability of interstitials at lower energy states would increase $E_a$ at low temperatures, while some molecular dynamics simulations\cite{lee_comparative_2014} argue that those atoms trapped at low energy sites hardly contribute to diffusion and hence $E_a$ would be lower at low temperatures. However, we believe that the measured $E_a$, which is a parameter averaged over all diffusing interstitials, is a characteristic of the potential energy surface mainly determined by the host structure (at least for low H concentrations where H-H interactions can be neglected). At low temperatures, the trapping of H interstitials at those low energy sites just implies more time is needed for them to sample the potential energy surface, which results in a decreased $D$ but not necessarily the change in $E_a$. \hlt{In other words, temperature change only changes $D$ of H interstitials, but not $E_a$, as long as the host structure (and hence the potential energy surface) is not changed.}

\begin{figure}
\includegraphics[width=3.3in]{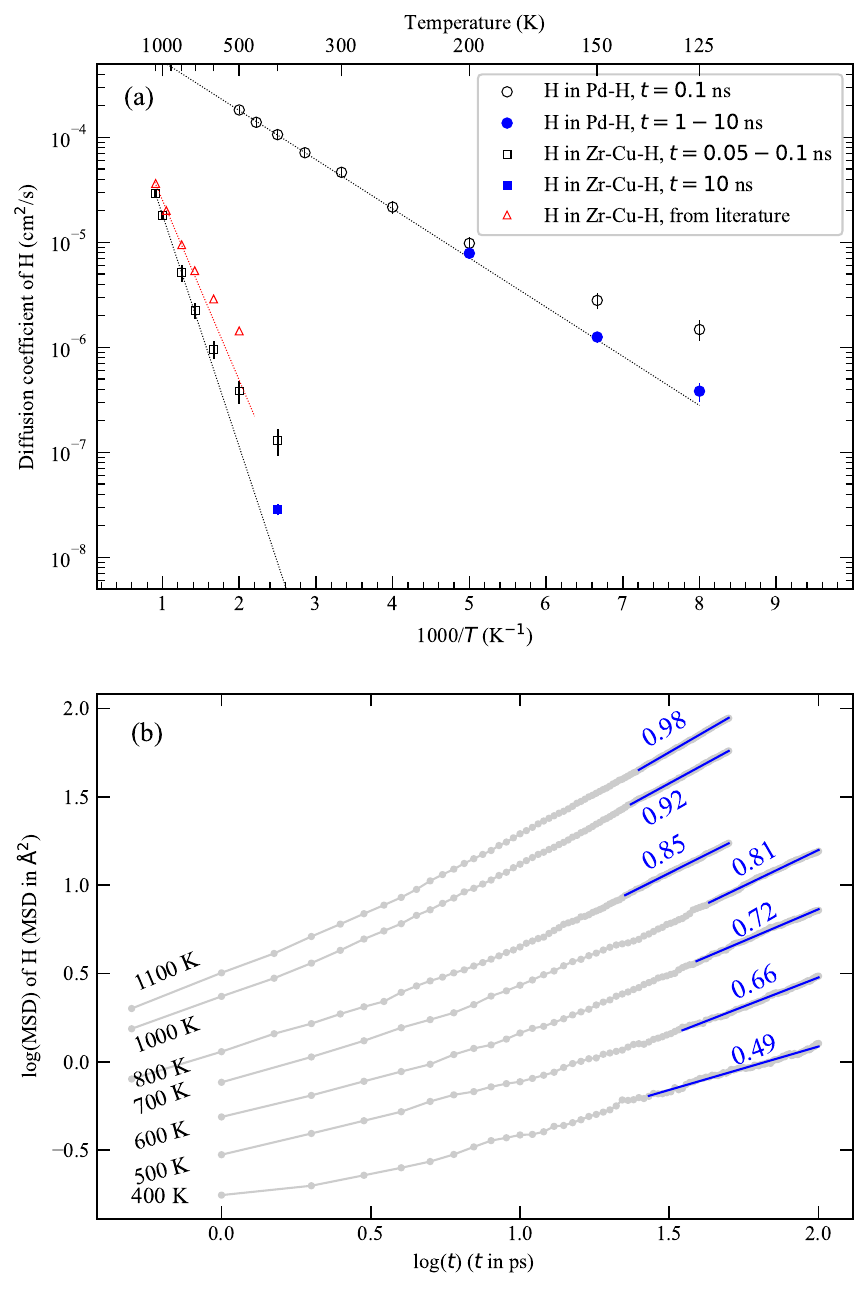}
\caption{(a) Computed diffusion coefficient. The triangular symbols are the values of $D$ replotted from reference \citenum{lee_comparative_2014}. The data difference between this work and reference \citenum{lee_comparative_2014} is probably due to different time spans used. The straight lines are for eye guide. \hlt{The blue solid symbols at relatively low temperatures indicate the improved accuracy from extended simulations.} (b) Mean squared displacements of H in Zr-Cu-H alloy. The top 50\% MSD (highlighted in blue) are linearly fitted, and the slopes are indicative of the convergence of $D$ calculations. For the calculation at 400 K for $t$=10 ns (not shown), the slope increases to 0.64.}
\label{fig:d}
\end{figure}

To test our above arguments, we computed MSDs of hydrogen in amorphous Pd-H and Zr-Cu-H alloys (without fixing the host atoms) at various temperatures. The time span for MSD calculations was 50 ps for temperatures $T\geq$800 K and 100 ps for lower temperatures. The time span chosen here is comparable to that (30-90 ps) used for previous simulations of Zr-Cu-H alloys\cite{lee_comparative_2014}. The obtained data for diffusion coefficient based on Equation \ref{eq:Einstein} are shown in Table \ref{tab:d} and plotted in Fig. \ref{fig:d}(a). As can be seen, the Arrhenius plots for both alloys exhibit a positive curvature, similar to the previous molecular dynamics results\cite{lee_comparative_2014}.

\begin{figure}[t]
\includegraphics[width=3.3in]{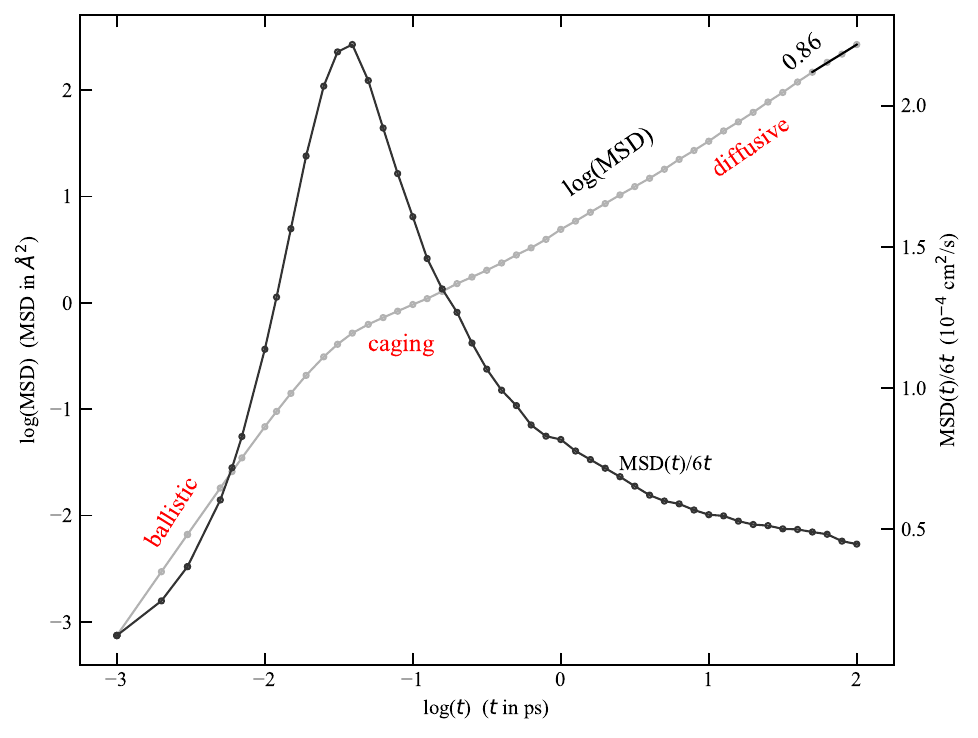}
\caption{Mean squared displacement of H in Pd-H at 300 K. Three regimes can be identified. The log(MSD)/log($t$) slope of the top 50\% MSD is 0.86.}
\label{fig:pdmsd}
\end{figure}

However, it turns out that these curved Arrhenius plots are due to the systematic errors resulting from the too short time span used for the calculations. As shown in Fig. \ref{fig:pdmsd}, the MSD of atoms as a function of time can be divided into three regimes \cite{donati_spatial_1999}. During the `ballistic' regime, atoms move without colliding with other atoms, and MSD is proportional to $t^2$. In the `caging' regime, atoms collide with their neighbours and move forth and back, and MSD is roughly a constant. The timescale for the caging regime is temperature- and material-dependent. When the system enters the deep `diffusive' regime, MSD is proportional to $t$ and the Einstein relation applies. In the log scales, the ballistic regime has a slope of 2 and the deep diffusive regime has a slope of 1. Corresponding to these three stages, the parameter MSD/6$t$=$\langle r^2(t) \rangle/6t$ increases to a maximum at the end of the ballistic regime and then converges towards the diffusion coefficient defined by Equation \ref{eq:Einstein}. Shown as an example in Fig. \ref{fig:d}(b), the slopes of log(MSD)/log($t$) for the Zr-Cu-H alloy are far from 1 at low temperatures, implying the values of $D$ are far from convergence. For a remedy, we extended the simulation time for Pd-H for temperatures below 250 K, and the new Arrhenius plot is well linearly fitted. As shown in the figure, extension of simulation time for the Zr-Cu-H alloy also improves the accuracy. We finally note that, since amorphous phases lack the structural periodicity as in crystals, the convergence of $D$ with respect to MSD for amorphous materials is generally slower than that for crystals and hence larger MSDs or longer simulation time is often necessary for accurate computation of $D$.

The above results indicate that hydrogen diffusion in amorphous metals follows the Arrhenius relation, which is consistent with the experiments as mentioned earlier. The remaining question is why non-Arrhenian diffusion is sometimes observed. It is key to note that in this scenario the Arrhenius curves break into several {\it{linear segments}} \cite{bowman_nmr-studies_1981, bowman_proton_1982, bowman_hydrogen_1983}, exhibiting abrupt changes in $E_a$, instead of a smooth $E_a$ variation over a large temperature range as predicted by previous simulations\cite{kirchheim_modeling_1985, lee_comparative_2014}. Such abrupt changes often imply new diffusion mechanisms and/or structural changes of the alloys. For example, the change in $E_a$ for amorphous TiCuH$_{1.4}$ \cite{bowman_proton_1982} and Zr$_2$PdH$_{2.88}$ \cite{bowman_hydrogen_1983}
at $\sim$210 K is possibly due to quantum tunneling in view that similar transitions in hydrogenated  ZrCr$_2$ crystals were observed between 150 and 200 K and attributed to quantum tunneling\cite{renz_pulsed-field-gradient_1994}. The decomposition of amorphous TiCuH$_x$ ($x\sim$1.3) at $\sim$420 K was found to cause an abrupt drop of hydrogen diffusion coefficient \cite{bowman_nmr-studies_1981}. For other $E_a$ changes in amorphous alloys \cite{bowman_nmr-studies_1981, bowman_proton_1982, bowman_hydrogen_1983}, no associated phase transformations, such as crystallization, were reported. However, these studies were carried out in decades ago when our knowledge of amorphous structures was limited. Recent studies reveal amorphous-amorphous phase transitions \cite{lan_hidden_2017, liu_evidence_2007} in amorphous alloys, and such transitions may change the medium-range packing of atomic clusters and hence impact on the diffusion of hydrogen. \hlt{From the perspective of potential energy landscape, upon phase transition the system moves from one subset of the potential energy landscape to another subset and hence exhibits different energy characteristics or diffusion activation energy. This is not the case for the systems studied here.}  

\section{Summary}
In this work we studied the diffusion of hydrogen in amorphous PdH$_{0.05}$ and (ZrCu)H$_{0.1}$ alloys based on molecular dynamics simulations. Our simulations confirm the distribution of location-dependent hydrogen interstitial energies and the influence of temperature on the distribution of hydrogen interstitials. \hlt{It was found that hydrogen interstitials prefer the Voronoi vertices of the host atoms. For the Pd-H system, 90\% H atoms were found to be within 0.3 $ $\AA$ $ from the vertices and, as temperature increases, the potential energy of H atoms fluctuates in a wider range with the most probable potential energy also increasing.} In a local scale, those hydrogen interstitials of higher potential energy are more mobile than those of lower energy. However, in a global scale the activation energy of hydrogen diffusion reflects the features of the potential energy landscape that depends on the material structure. Consistently, we found Arrhenian diffusion behaviour for hydrogen in amorphous alloys, in contrast to previous theoretical studies which predict non-Arrhenian behaviour. \hlt{While only a single composition for each alloy system was studied in this work, we believe the conclusion for Arrhenian diffusion of hydrogen is general since the potential energy landscape (which determines the diffusion activation energy) is fixed for a given composition. A lesson from this work is that we should be cautious about the convergence when computing diffusion coefficient of amorphous structures based on molecular dynamics. Because the potential energy landscape of amorphous structures is generally more complicated than that of crystalline structures, usually a longer time is needed for sampling the landscape and obtaining the correct activation energies.} We also discussed the experimental segmented Arrhenian behaviour of hydrogen diffusion within the framework of quantum tunneling and amorphous-amorphous phase transformations.

\end{document}